\documentclass[conference]{IEEEtran}
\IEEEoverridecommandlockouts
\usepackage{cite}
\usepackage{amsmath,amssymb,amsfonts}
\usepackage{algorithmic}
\usepackage{graphicx}
\usepackage{textcomp}
\usepackage{xcolor}
\usepackage{bm}
\usepackage{multirow}
\usepackage{color}
\usepackage{subfigure}

\begin{document}

\title{Phase Code Discovery for Pulse Compression Radar: A Genetic Algorithm Approach
}
\author{
Xinyan~Xie,
Runxin~Zhang,
Yulin~Shao,~\IEEEmembership{Member,~IEEE},
Lu Lu,~\IEEEmembership{Member,~IEEE}

\thanks{X. Xie, R. Zhang, and L. Lu are with the University of Chinese Academy of Sciences, and the Key Laboratory of Space Utilization, Chinese Academy of Sciences, Beijing 100094, China (emails: \{xiexinyan20, zhangrunxin20\}@mails.ucas.ac.cn, lulu@csu.ac.cn).

Y. Shao is with the Department of Electrical and Electronic Engineering, Imperial College London, London SW7 2AZ, U.K. (e-mail: y.shao@imperial.ac.uk).
}
}

\maketitle

\begin{abstract}
Discovering sequences with desired properties has long been an interesting intellectual pursuit.
In pulse compression radar (PCR), discovering phase codes with low aperiodic autocorrelations is essential for a good estimation performance. The design of phase code, however, is mathematically non-trivial as the aperiodic autocorrelation properties of a sequence are intractable to characterize.
In this paper, we put forth a genetic algorithm (GA) approach to discover new phase codes for PCR with the mismatched filter (MMF) receiver.
The developed GA, dubbed GASeq, discovers better phase codes than the state of the art. At a code length of $59$, the sequence discovered by GASeq achieves a signal-to-clutter ratio (SCR) of $50.84$, while the best-known sequence has an SCR of $45.16$.
In addition, the efficiency and scalability of GASeq enable us to search phase codes with a longer code length, which thwarts existing deep learning-based approaches. At a code length of $100$, the best phase code discovered by GASeq exhibit an SCR of $63.23$.
\end{abstract}


\section{Introduction}\label{sec:I}
\begin{figure}[t]
\centering
\includegraphics[width=1\linewidth,scale=1.00]{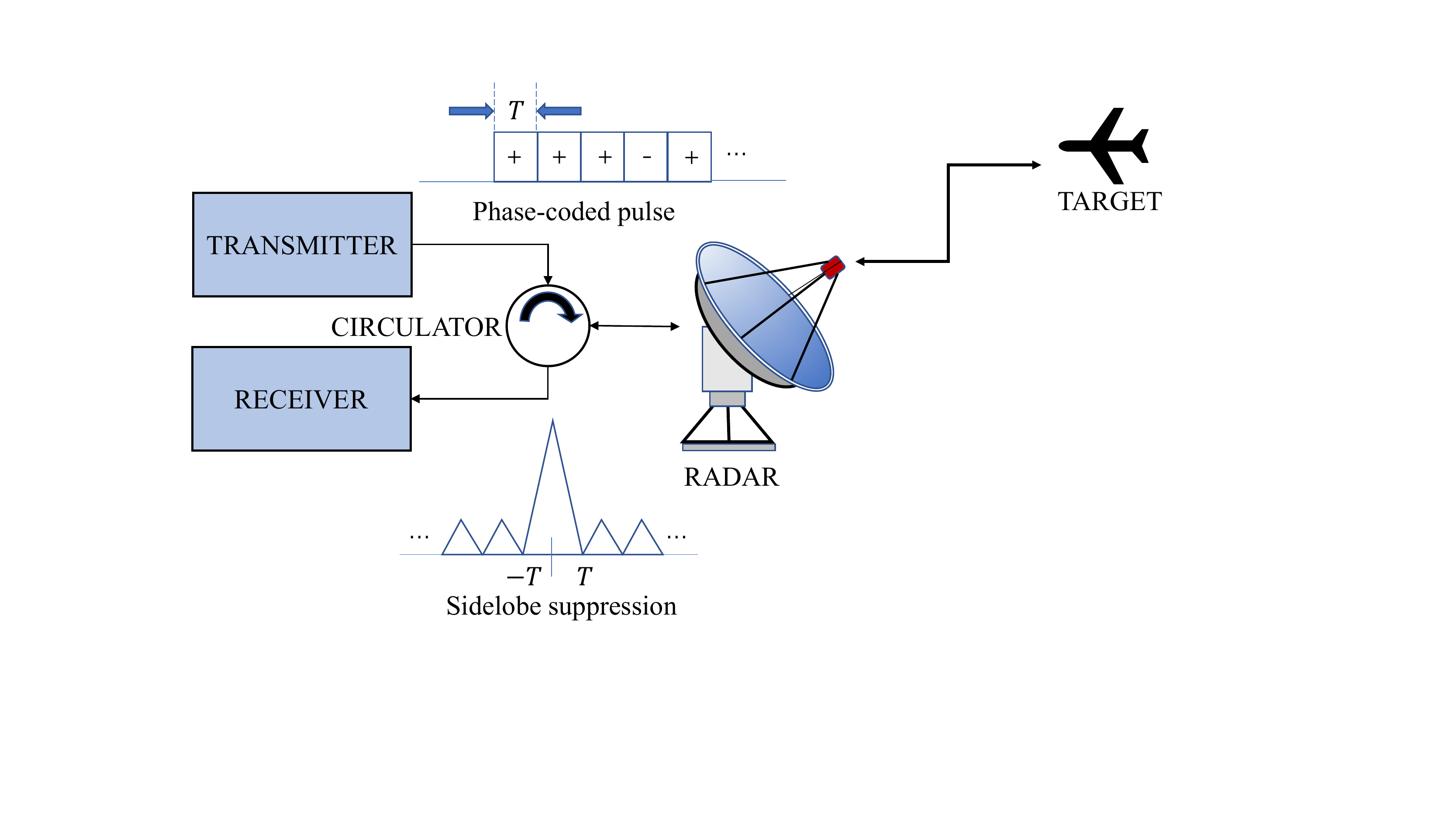}	
\caption{Schematic diagram of PCR: a phase-coded pulse is emitted; the design objective of the phase code and receiver is sidelobe suppression.}
\label{fig:PCR}
\end{figure}

Pulse compression radar (PCR) is a class of radar that solves the detection range and resolution trade-off in classical radar systems \cite{PCR}.
As illustrated in Fig.~\ref{fig:PCR}, the principles of PCR are 1) modulating the transmitted pulse by a sequence of phase codes, and 2) matched filtering (MF) the received signal by the same phase code, which is known as the MF receiver \cite{meritfactor}.
In so doing, PCR has both a long detection range, because the power of the whole pulse is collected by MF, and a high detection resolution, because the modulated pulse exhibits a large bandwidth -- the detection resolution of radar is proportional to the bandwidth of the transmitted pulse \cite{rangeresolution}.

When developing a PCR system, the main design objective is sidelobe suppression by carefully crafted phase codes and the receiver structure. High sidelobes are detrimental to PCR as they increase both the miss-detection rate and the false alarm rate \cite{survey_radar}.
In the literature, there are two main receiver structures: the MF receiver \cite{Legendre,xLastovkaMF,survey_MF} and the mismatched filter (MMF) receiver \cite{MMF_1990,MRLS,IVreciever}.

Most research efforts on PCR have been devoted to the design of binary phase code with the MF receiver \cite{survey_MF}. In actuality, this problem corresponds to the famous merit factor problem in complex analysis, that is, discovering the binary sequence with the smallest aperiodic autocorrelations.
The main techniques are theoretical approaches \cite{Legendre}, constrained search \cite{xLastovkaMF}, exhaustive computation, and stochastic search. We refer readers to the excellent survey \cite{survey_MF} for more detailed explanations.
On the other hand, as far as the PCR estimation performance is concerned, the other line of receiver design, i.e., the MMF receiver, is more promising.
Unlike the MF receiver that cross-correlates the received pulse by the same binary phase code, the MMF receiver utilizes a real sequence to collect the received power \cite{MMF_1990,IVreciever} and optimizes the real sequence such that the signal-to-clutter ratio (SCR) -- which is also known as the signal-to-interference ratio (SIR) -- is maximized, thereby achieving remarkable performance gains over the MF receiver \cite{IVreciever}. The phase code design for the MMF receiver, however, is relatively few because of the lack of mathematical instruments.

Motivated by data-driven approaches such as deep learning (DL) \cite{AlphaGo,Significant2020,hpgan}, recent works \cite{AlphaSeq,hpgan} revisit the phase-code design problem for PCR with the MMF receiver and discover good phase codes with high SCR. The authors in \cite{AlphaSeq} proposed a deep reinforcement learning (DRL) approach, named AlphaSeq, to search for good phase codes. In AlphaSeq, the phase code design problem is described as a game and the SCR of the discovered phase code is defined as the reward. Over the course of DRL, the algorithm learns to discover better and better phase codes with increasingly higher SCR.
On the other hand, HpGAN \cite{hpgan} utilizes the DL-based generative model, i.e., generative adversarial network (GAN), to generate new phase codes.

In this paper, we put forth a genetic algorithm (GA) approach, dubbed GASeq, to discover phase codes with low aperiodic autocorrelations for PCR with the MMF receiver. Compared with DL-based schemes, our approach discovers better phase code with higher SCR. Specifically, at a length of $59$, the best phase code discovered by GASeq achieves an SCR of $50.84$, while the best sequences found by AlphaSeq and HpGAN have an SCR of $33.45$ and $45.16$, respectively.
In addition, our approach is much more efficient than DL-based approaches. Compare with AlphaSeq, for example, GASeq visits $166$ times fewer states in order for the algorithm to converge and the discovered sequence is much better. Thanks to the high efficiency and scalability, we extend GASeq to discover longer phase codes. At a length of $100$, the discover sequence achieves an SCR of $63.23$.

The remainder of this paper is organized as follows.
Section \ref{sec:II} describes the system model for PCR and introduces the MMF receiver.
Section \ref{sec:III} details the proposed GA approach, i.e., GASeq.
Section \ref{sec:IV} evaluates the performance of GASeq benchmarked against prior arts.
Section \ref{sec:V} concludes this paper.

{\it Related work on GA} -- Inspired by Charles Darwin's ``natural selection'', GA was first introduced by John Holland \cite{GA_1992} in 1975 as a metaheuristic approach to solve optimization problems \cite{GA_review1}.
GA has been applied to a variety of disciplines \cite{GA_review2}, such as network routing protocol, image processing, data mining, neural networks, to name a few.

GA has also been used in radar systems.
In \cite{phasearray}, for example, the authors proposed a GA to divide the phase array of radar into the subarrays in order to reduce the hardware cost.
Ref. \cite{MIMOradar} applied GA to a multiple-input and multiple-output (MIMO) radar to find the best locations for transmit and receive antenna arrays.
There have also been studies in GA for PCR with the MF receiver.
In \cite{GAPSL}, the authors enhanced the GA by a local search scheme and find phase codes with good peak sidelobe levels.
In \cite{MGADF}, the authors combined the global minimum convergence property of GA with the fast convergence rate of hamming scanning algorithm, and discovered phase codes with good discriminating factors.

\section{System model}\label{sec:II}
We consider a phase-coded pulse compression radar system, where the transmitted pulse is modulated by a sequence of binary codes $\bm{s}=\{s_n\in\{+1,-1\}:n=0,1,...,N-1\}$, and $+1$ and $-1$ correspond to phases $0$ and $\pi$, respectively.
The received signal, on the other hand, is a sum of many echoes from various range bins with different amplitudes and delays.
In particular, we denote by $h_0$ the radar cross section (RCS) of the range bin of interest.
The received discrete-time model is given by
\begin{equation}\label{eq:y}
\bm{y}=h_0\bm{s}+\sum_{i=1-N, i\neq 0}^{N-1}h_i\bm{J}_i\bm{s}+\bm{w},
\end{equation}
where $\{h_i:i=1-N,2-N,...,N-2,N-1,i\neq 0\}$ denotes the RCS of interfering range bins; $\bm{w}$ is the white Gaussian noise vector; $\bm{J}_i$, $\forall i$, denote $N\times N$ shift matrices that capture the propagation time needed for the clutters to reach the radar receiver. In particular, $\bm{J}_i$ can be written as 
\begin{equation}\label{matrix}
\bm{J}_i \triangleq
\bordermatrix{%
       & 1         & \cdots  &i          & \cdots     & N\cr
1      & 0         &         & 1         & \cdots     & 0\cr
       &           & 0       &           & 1          &\cr
\vdots & \vdots    &         & 0         &            & 1\cr
       &           &         &           & 0          &\cr 
N      & 0         &         &           &            & 0\cr
},
\end{equation}
and $\bm{J}_i = \bm{J}_{-i}^\top$.

Given the received signal $\bm{y}$, our goal is to estimate $h_0$, the SCR of the range bin of interest. To this end, we cross-correlate the received sequence $\bm{y}$ by a real vector $\bm{x}$, yielding
\begin{equation}\label{MMFreceiver}
\bm{x}^\top\bm{y}=h_0\bm{x}^\top\bm{s}+\sum_{i=1-N, n\neq 0}^{N-1}h_i\bm{x}^\top\bm{J}_i\bm{s},
\end{equation}
where $\bm{w}$ is omitted since the noise term is often dominated by interference.

Note that the RCS of different range bins $\{h_i\}$ are unknown to the radar receiver.
The sequence discovery problem in PCR is then discovering the optimal pair of sequences $(\bm{s,x})$ such that the signal-to-clutter ratio (SCR) $\gamma$ is maximized, where
\begin{equation}\label{MMF}
\gamma=\frac{(\bm{x}^\top\bm{s})^2}{\sum_{i=1-N, i\neq 0}^{N-1}(\bm{x}^\top\bm{J}_i\bm{s})^2}.
\end{equation}

Moreover, let $\bm{R}\triangleq\sum_{i=1-N, i\neq 0}^{N-1}\bm{J}_i\bm{s}\bm{s}^\top\bm{J}_i^\top$, $\gamma$ can further be simplified as \cite{IVreciever}:
\begin{eqnarray}\label{MMF1}
\gamma \hspace{-0.2cm}&=&\hspace{-0.2cm}\frac{(\bm{x}^\top\bm{s})^2}{\bm{x}^\top\bm{R}\bm{x}}
=\frac{(\bm{x}^\top\bm{R}^{\frac{1}{2}}\bm{R}^{-\frac{1}{2}}\bm{s})^2}{\bm{x}^\top\bm{R}\bm{x}}
\overset{(a)}{\leq}
\frac{(\bm{x}^\top\bm{R}\bm{x})(\bm{s}^\top\bm{R}^{-1}\bm{s})}{\bm{x}^\top\bm{R}\bm{x}} \nonumber\\
\hspace{-0.2cm}&=&\hspace{-0.2cm}\bm{s}^\top\bm{R}^{-1}\bm{s},
\end{eqnarray}
where (a) follows from the Cauchy-Schwartz inequality. The equality holds when $\bm{R}^{\frac{1}{2}}\bm{x}=\bm{R}^{-\frac{1}{2}}\bm{s}$. Thus, for a given $\bm{s}$, the optimal $\bm{x}^*$ that maximizes $\gamma$ is  $\bm{x}^*=\bm{R}^{-1}\bm{s}$.
The sequence discovery problem in SCR can be refined as
\begin{equation}\label{MMFoptimal}
\bm{s}^*=\arg\max_{\bm{s}\in\{+1,-1\}^N} \bm{s}^\top\bm{R}^{-1}\bm{s}.
\end{equation}

The combinatorial optimization problem in \eqref{MMFoptimal} is non-trivial to solve analytically. Thus, prior work often resorts to algorithmic solutions such as exhaustive search \cite{IVreciever}, heuristic search \cite{xLastovkaMF}, and learning-based algorithms \cite{AlphaSeq,hpgan}. In this paper, we put forth a GA-based solution to discover better phase codes $\bm{s}$ for SCR.

\textbf{Remark:} {\it As prior works \cite{IVreciever,AlphaSeq,hpgan}, this paper does not consider fractional misalignment among echoes. If fractional misalignment is further considered, interested readers may refer to \cite{misalignedOAC,PNC,BayesianOAC} for signal processing techniques to process the received signal. In particular, \cite{PNC} deals with discrete source (e.g., binary phase codes), while \cite{misalignedOAC,BayesianOAC} deal with continuous source (e.g., polyphase codes).}

\section{GASeq: A Genetic Algorithm Approach}\label{sec:III}
GA algorithms solve combinatorial optimization problems by a natural selection process that mimics biological evolution: in the first iteration, it generates a population of phase codes as candidates; in subsequent iterations, it repeatedly improves the candidates based on the population of the last iteration such that the candidates evolve towards the optimal solution. In this section, we detail our design of the GASeq for phase-code discovery. 
\begin{figure}[t]  
\centering  
\includegraphics[width=\linewidth,scale=1.00]{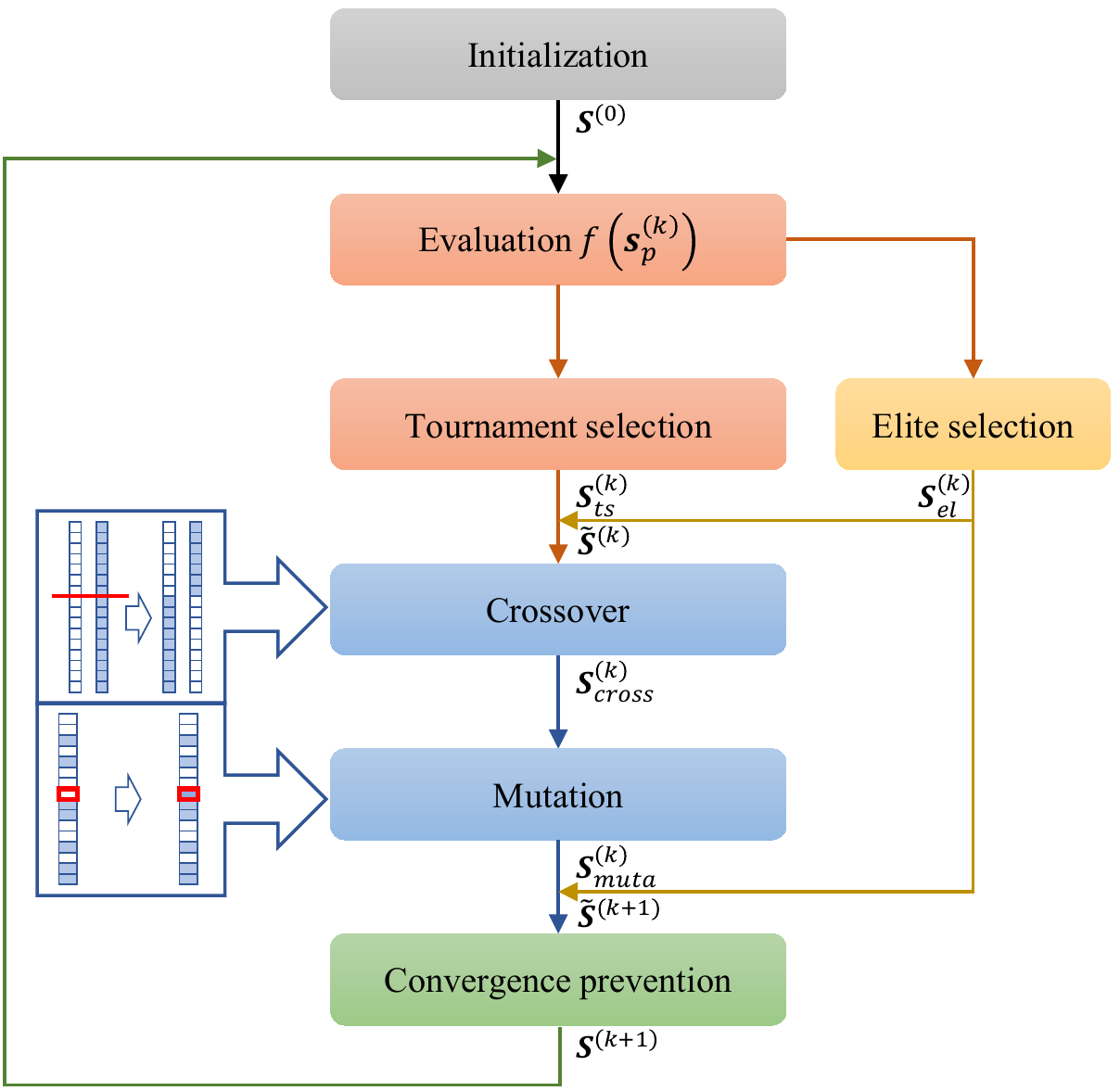}
\caption{The signal flow of GASeq for phase code discovery in PCR.}
\label{GA}
\end{figure}

To start with, we outline the signal flow of GASeq in Fig.~\ref{GA}.

\textbf{Initialization}. The initialization phase generates a population of phase codes as initial candidates. Let the population size be $P$, the initialized population can be written as an $N\times P$ matrix:
\begin{equation}\label{initial}
\bm{S}^{(0)}=\left[ \bm{s}^{(0)}_1,\bm{s}^{(0)}_2,\cdots,\bm{s}^{(0)}_{P} \right],
\end{equation}
where each column $\bm{s}^{(k)}_{p}\in\{+1,-1\}^N$ is a phase code of length $N$; the superscript $k=0,1,2,...,K$ denotes the index of iteration (there are $K+1$ iterations and the initialized population is the $0$-th iteration); and the subscript $p=1,2,...,P$ denotes the index of the phase code in the population.

In general, the elements of $\bm{S}^{(0)}$ are sampled uniformly from $\{+1,-1\}$. 
Another idea is incorporating existing phase codes discovered by AlphaSeq \cite{AlphaSeq} and HpGAN \cite{hpgan} into $\bm{S}^{(0)}$ to generate more promising offspring and speed up the convergence. We will compare different initialization schemes in Section \ref{sec:IV}.

\textbf{Evaluation}. The evaluation process computes the ``fitness'' of individual phase code in the current population, where fitness reflects how good a phase code is in terms of \eqref{MMFoptimal}, and can be computed by $f(\bm{s}^{(k)}_p)\triangleq(\bm{s}^{(k)}_p)^\top\bm{R}^{-1}\bm{s}^{(k)}_p$, $\forall \bm{s}^{(k)}_p\in\bm{S}^{(k)}$.


Based on the current population $\bm{S}^{(k)}$ and the evaluated fitness, we next determine the next population $\bm{S}^{(k+1)}$.

\textbf{Elite selection}.
To goal of selection is to retain superior candidates and eliminate inferior candidates from the last population according to their fitness.
In order to retain the best candidates of $\bm{S}^{(k)}$, we first perform elite selection to identify $E$ phases codes in $\bm{S}^{(k)}$ with the largest fitness. Let us denote the elite codes by $\bm{S}_{el}^{(k)}\in\mathcal{R}^{N\times E}$, which is a submatrix of $\bm{S}^{(k)}$.
The elite codes will be directly assigned to $\bm{S}^{(k+1)}$, while the other $P-E$ vacancies in $\bm{S}^{(k+1)}$ are determined by tournament selection, as detailed below.

\textbf{Tournament selection}.
In this operation, we perform $P-E$ tournaments \cite{tour} to select $P-E$ candidates to form a new matrix $\bm{S}_{ts}^{(k)}\subset\bm{S}^{(k)}$.
In each tournament, we randomly select $M$ phase codes without replacement from $\bm{S}^{(k)}$ (including the elites) and compare their fitnesses. The winner, i.e., the phase code with the maximal fitness among the chosen $M$ phase codes, goes to $\bm{S}_{ts}^{(k)}$.

For the phase code with the $i$-th largest fitness in $\bm{S}^{(k)}$, the probability that it wins a tournament is 
\begin{equation}
p_i=\frac{M\cdot \left(
\begin{matrix} P-i \\ M-1 \end{matrix} \right)}{\left(
\begin{matrix} P \\ M \end{matrix} \right)}=\frac{M^2\cdot (P-i)!(P-M)!}{P!(P-i-M+1)!}.
\label{eq:probablity}
\end{equation}
Therefore, the probability that it goes to  $\bm{S}_{ts}^{(k)}$ after tournament selection is $1-(1-p_i)^{P-E}$.

\textbf{Crossover and mutation}.
Crossover refers to the operations of replacing and recombining the parts of a pair of candidates (i.e., parents) to generate a new candidate (i.e., child).
To start with, we combine the tournament-selected phase codes
$\bm{S}_{ts}^{(k)}$ with the elites $\bm{S}_{el}^{(k)}$, yielding $\widetilde{\bm{S}}^{(k)}=\left[\bm{S}_{ts}^{(k)},\bm{S}_{el}^{(k)}\right]$.

With crossover, we first randomly select two phase codes $\bm{s}^{(k)}_i$, $\bm{s}^{(k)}_j$, $i,j\in \left\{1,2,\cdots,P\right\}$ from $\widetilde{\bm{S}}^{(k)}$.
Then, we split both $\bm{s}_i^{(k)}$ and $\bm{s}_j^{(k)}$ into two parts at a random point and form a new phase code by concatenating the first part of  $\bm{s}_i^{(k)}$  and the second part of  $\bm{s}_j^{(k)}$.
For example, if $\bm{s}_i^{(k)}=[1,-1,1,-1,1,-1,1]^\top$, $\bm{s}_j^{(k)}=[-1,-1,-1,1,1,1,1]^\top$, and we split after the third symbol, the newly constructed phase code is given by $\bm{s}_{\text{new}}^{(k)}=[1,-1,1,1,1,1,1]^\top$.
The crossover operation will be performed $P-E$ times, after which we obtain $P-E$ new phase codes, forming a new set $\bm{S}_{\text{cross}}^{(k)}\in\mathcal{R}^{N\times(P-E)}$.

Next, we randomly mutate the elements of $\bm{S}_{\text{cross}}^{(k)}$.
The purpose is to explore the search space and introduce more diversity to the algorithm.
To be more specific, for each phase code in $\bm{S}_{\text{cross}}^{(k)}$, we mutate one element (from $-1$ to $1$ or from $1$ to $-1$) with probability $p_{\text{muta}}$ ($0\leq p_{\text{muta}}\leq1$).
Denoting by $\bm{S}_{\text{muta}}^{(k)}$ the mutated $\bm{S}_{\text{cross}}^{(k)}$, we concatenate it with the elite codes of $\bm{S}^{(k)}$, yielding $\widetilde{\bm{S}}^{(k+1)}=\left[\bm{S}_{\text{muta}}^{(k)},\bm{S}_{\text{el}}^{(k)}\right]$.

\textbf{Early-convergence prevention}. The last step of one iteration is early-convergence prevention. Notice that in the above operations, a phase code in $\bm{S}^{(k)}$ can appear in $\widetilde{\bm{S}}^{(k+1)}$ multiple times, and the average number of appearance is proportional to the  fitness of the phase code. To prevent the algorithm from being dominated by these ``better'' phase codes and early stopping, we have to reduce the number of repeating phase codes in the new generation.

The early-convergence prevention operates in the following manner.
Suppose that a phase code $\bm{s}$ appears in $\widetilde{\bm{S}}^{(k+1)}$ for $G$ times. For each appearance, we will decide independently whether to keep or drop it.
First, the first appearance is kept to ensure that $\bm{s}$ appears at least once in $\bm{S}^{(k+1)}$.
Then, the rest appearances will be kept with probability $p_{\text{conv}}$, where $0\leq p_{\text{conv}}\leq 1$, and dropped from $\widetilde{\bm{S}}^{(k+1)}$ otherwise. After the above operations, we obtain a new population $\widetilde{\bm{S}}_{\text{left}}^{(k+1)}$ with the number of phase codes less than $P$.

Finally, the $(k+1)$-th generation of phase codes $\bm{S}^{(k+1)}$ is obtained by padding randomly initialized phase codes to   $\widetilde{\bm{S}}_{\text{left}}^{(k+1)}$ such that $\bm{S}^{(k+1)}\in\mathcal{R}^{N\times P}$.

\section{Numerical Experiments} \label{sec:IV}
Given the GASeq algorithm described in Section \ref{sec:III}, this section performs numeral experiments to discover new phase codes for pulse compression radar with the MMF receiver. 

\subsection{Experimental setup and baselines}
We adopt the same system setup as \cite{AlphaSeq,hpgan} and the goal is to discover a phase code of length $N=59$ that maximizes the SCR $\gamma$.
The default hyper-parameter setting of our GASeq is summarized in Table~\ref{tab:1}.
Specifically, the algorithm operates for $N_G$ generations and the population size is $P=10,000$. The elite population size is set to $E=2000$ and the number of competitors in each tournament is $M=5$. The mutation rate is $p_{\text{muta}}=0.3$ and the probability for convergence prevention is $p_{\text{conv}}=0.7$.

\begin{table}[t]
\caption{Hyperparameters of GASeq for Phase Code Discovery}
\begin{tabular}{|c|c|c|}
\hline
\multirow{8}{*}{GA} & \textbf{Parameters} & \textbf{Definitions}                         \\ 
\cline{2-3}& $N=59$& Length of the binary phase code\\
\cline{2-3}& $N_G=200$& Number of generations in GA \\ 
\cline{2-3} & $P=10000$ & Number of candidates in the population\\ 
\cline{2-3} & $E=2000$ & Number of elites\\
\cline{2-3}& $M=5$& Number of competitors in a tournament selection \\
\cline{2-3} & $p_{\text{muta}}=0.3$ & Mutation rate\\ 
\cline{2-3}& $p_{\text{conv}}=0.7$& Probability for convergence prevention\\ \hline
\end{tabular}
\label{tab:1}
\end{table}

There are three baselines: the Legendre sequence \cite{Legendre}, AlphaSeq \cite{AlphaSeq}, and HpGAN \cite{hpgan}. 
Let $N=59$, the Legendre sequence is given by
\begin{equation}
\bm{s}_\text{L}=
\begin{bmatrix}
\begin{smallmatrix}
+1 & +1 & -1 & +1 & +1 & +1 & -1 & +1 & -1 & +1\\ 
-1 & -1 & +1 & -1 & -1 & +1 & +1 & +1 & -1 & +1\\
+1 & +1 & +1 & -1 & -1 & +1 & +1 & +1 & +1 & +1\\
-1 & -1 & -1 & -1 & -1 & +1 & +1 & -1 & -1 & -1\\
-1 & +1 & -1 & -1 & -1 & +1 & +1 & -1 & +1 & +1\\
-1 & +1 & -1 & +1 & -1 & -1 & -1 & +1 & -1 &   \nonumber
\end{smallmatrix}
\end{bmatrix},
\end{equation}
and $\gamma(\bm{s}_\text{L})\approx 2.69$;
AlphaSeq is searched by DRL, giving
\begin{equation}
\textbf{s}_{\text{alpha}}=
\begin{bmatrix} 
\begin{smallmatrix}
+1 & +1 & +1 & +1 & +1 & +1 & +1 & +1 & +1 & +1\\ 
+1 & +1 & +1 & +1 & +1 & +1 & -1 & -1 & -1 & -1\\
-1 & -1 & -1 & -1 & +1 & +1 & +1 & -1 & -1 & +1\\
+1 & -1 & +1 & +1 & -1 & +1 & -1 & -1 & +1 & -1\\
+1 & -1 & +1 & -1 & -1 & +1 & -1 & +1 & -1 & +1\\
-1 & +1 & -1 & +1 & -1 & +1 & -1 & +1 & -1 &   \nonumber
\end{smallmatrix}
\end{bmatrix},
\end{equation}
and $\gamma(\bm{s}_{\text{alpha}})\approx 33.45$; HpGAN is searched by GAN, giving
\begin{equation}
\textbf{s}_{\text{HpGAN}}=
\begin{bmatrix} 
\begin{smallmatrix}
-1 & +1 & -1 & +1 & -1 & +1 & -1 & +1 & -1 & +1\\ 
-1 & +1 & -1 & +1 & -1 & +1 & +1 & -1 & +1 & -1\\
+1 & -1 & +1 & -1 & -1 & +1 & -1 & +1 & +1 & +1\\
-1 & +1 & +1 & -1 & -1 & -1 & -1 & +1 & +1 & +1\\
+1 & +1 & +1 & -1 & -1 & -1 & -1 & -1 & -1 & -1\\
-1 & -1 & -1 & -1 & -1 & -1 & -1 & -1 & -1 &   \nonumber
\end{smallmatrix}
\end{bmatrix},
\end{equation}
and $\gamma(\bm{s}_{\text{HpGAN}})\approx 45.16$.

\subsection{Main results}
\begin{figure}[t]
\centering
\includegraphics[width=0.8\linewidth,scale=1.00]{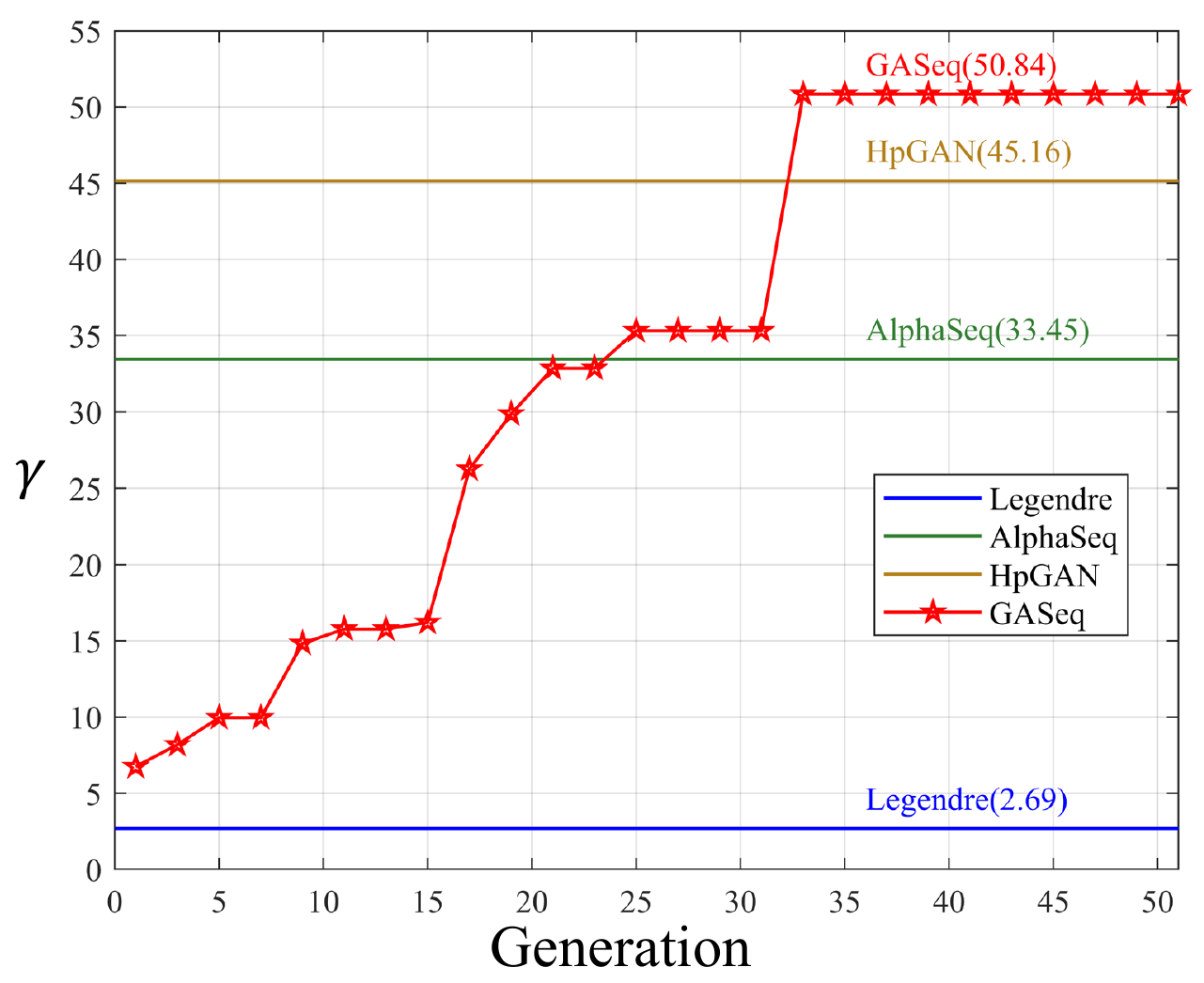}	
\caption{GASeq to discover a phase-coded sequence for pulse compression radar versus $\bm{s}_\text{L}$,$\bm{s}_{\text{alpha}}$, $\bm{s}_{\text{HpGAN}}$.}
\label{fig:sim1a}
\end{figure}

\begin{figure}[t]
\centering
\includegraphics[width=0.8\linewidth,scale=1.00]{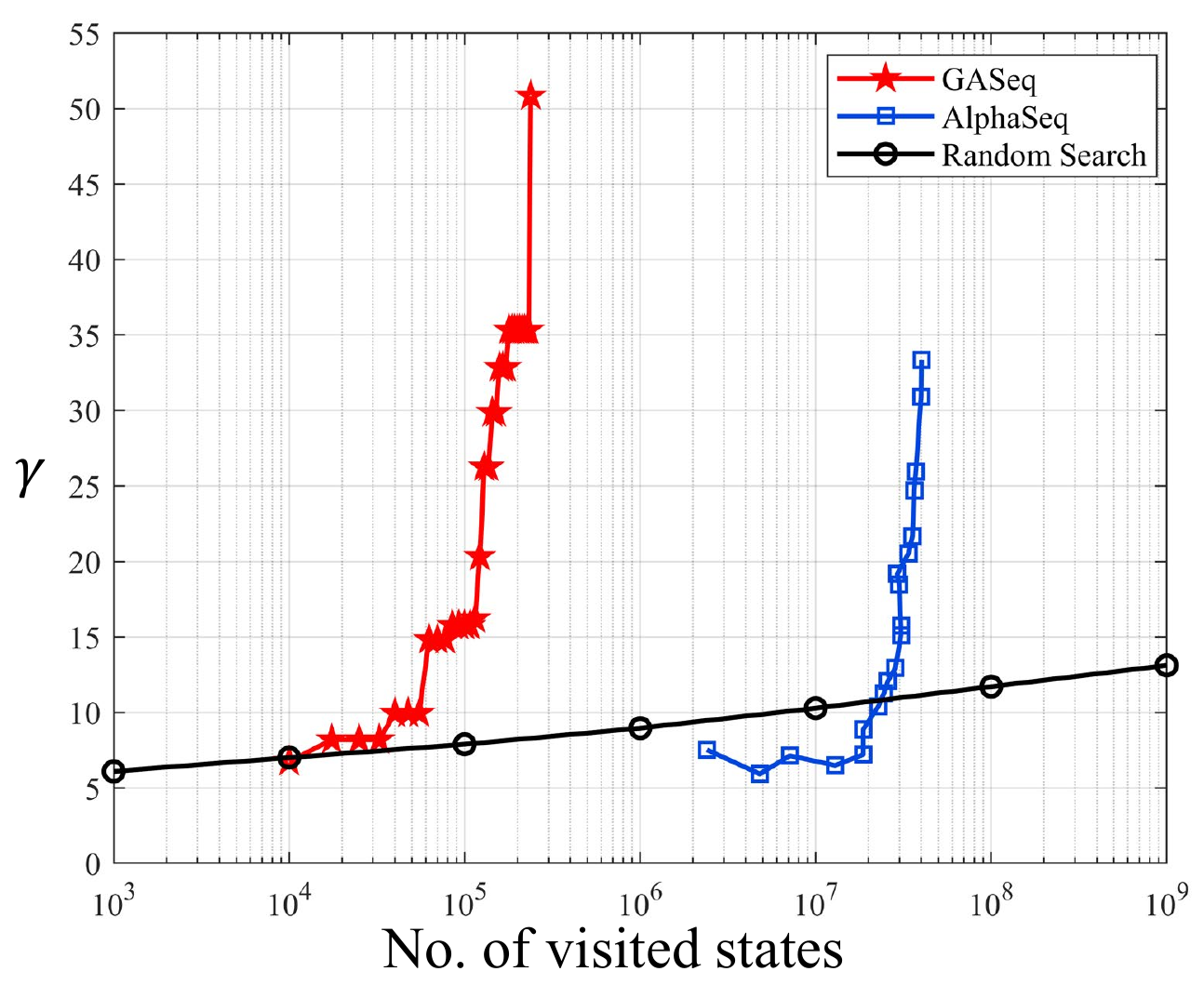}	
\caption{Searching capability comparison of GA, AlphaSeq, and random search.}
\label{fig:sim1b}
\end{figure}
To ease reading, we first summarize our main results in this subsection.
Given the hyperparameters in Table \ref{tab:1}, Fig.~\ref{fig:sim1a} presents the improvement of $\gamma$ over the course of evolution. As can be seen, after $32$ generations, GASeq converges to a phase code
\begin{equation}
\bm{s}_{\text{GA}}=
\begin{bmatrix}
\begin{smallmatrix}
+1 & +1 & +1 & +1 & +1 & +1 & +1 & +1 & +1 & +1\\ 
+1 & +1 & +1 & +1 & +1 & +1 & +1 & -1 & -1 & -1\\
-1 & -1 & +1 & +1 & +1 & -1 & -1 & +1 & +1 & +1\\
-1 & +1 & +1 & -1 & -1 & +1 & -1 & -1 & +1 & -1\\
+1 & -1 & -1 & +1 & -1 & +1 & -1 & +1 & -1 & +1\\
-1 & +1 & -1 & +1 & -1 & +1 & -1 & +1 & -1 &   \nonumber
\end{smallmatrix}
\end{bmatrix},
\end{equation}
the SCR of which is $\gamma(\bm{s}_{\text{GA}})=50.84$. Compared with $\bm{s}_\text{L}$, $\bm{s}_{\text{alpha}}$, and $\bm{s}_{\text{HpGAN}}$, $\bm{s}_{\text{GA}}$ improves the SCR by $48.15$, $17.39$ and $5.68$, respectively.

In addition, GASeq is a much more efficient search algorithm than prior arts. Fig.~\ref{fig:sim1b} compares the evolution of $\gamma$ as a function of the number of visited states for GA, DRL, and random search. As shown, GA discovers $\bm{s}_{\text{GA}}$ after visiting only $2.4\times 10^{5}$ states. At such a small number of visited states, the SCRs of the sequences discovered by AlphaSeq and random search are less than $10$.

\begin{figure}[t]
\centering
\includegraphics[width=0.8\linewidth,scale=1.00]{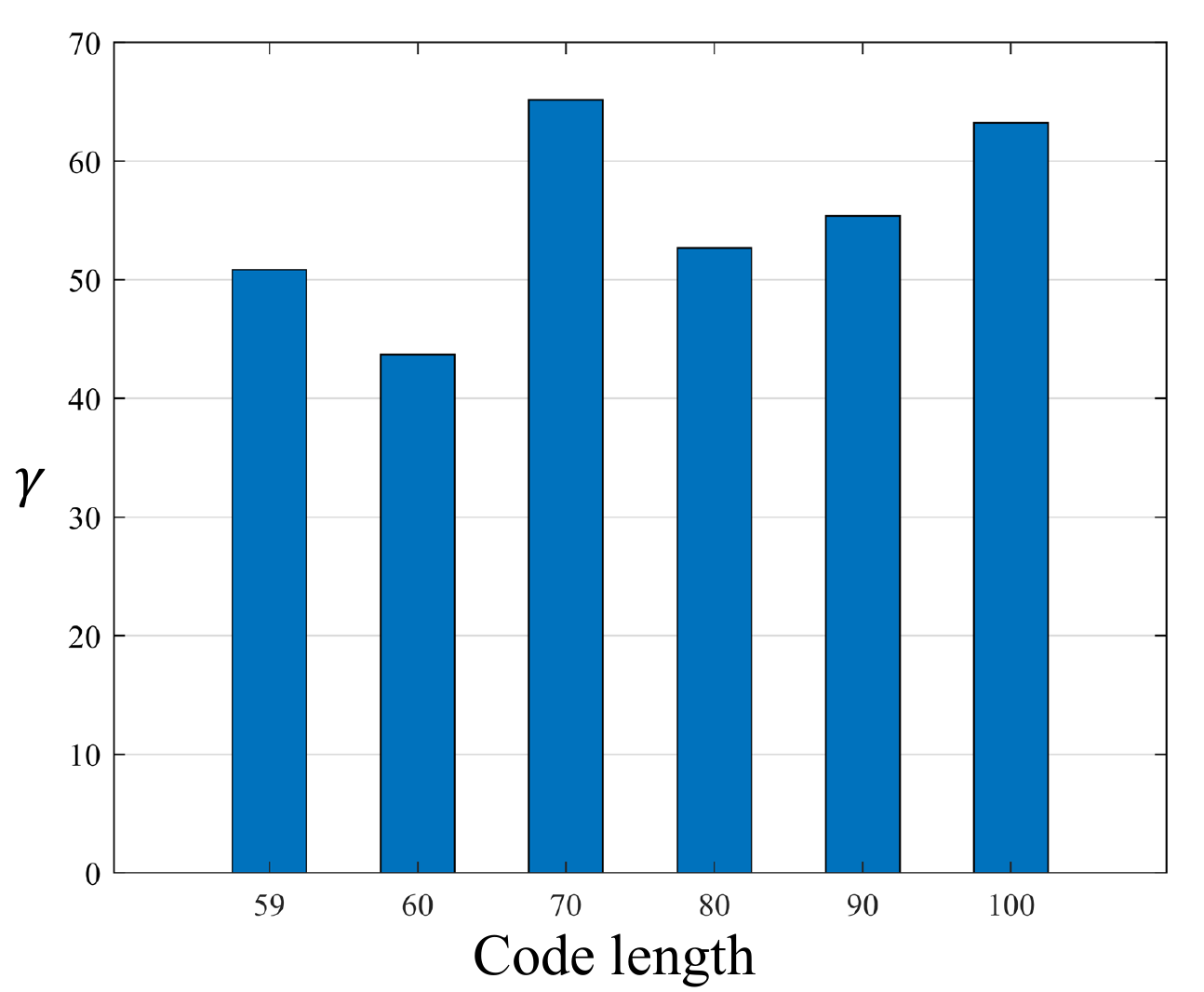}	
\caption{Impact of the different length of the phase codes $N$ on the performance of GASeq.}
\label{fig:change_N}
\end{figure}

Thanks to the efficiency and scalability of GASeq, we further apply it to discover longer phase codes for $N\in[60, 100]$ and present the achieved SCR in Fig.~\ref{fig:change_N}.
At a code length of $N=100$, for example, the best discovered sequence has an SCR of $63.23$. We emphasize that there is no proportional relationship between SCR and the code length -- as can be seen from \eqref{MMF}, with the increase in $N$, both the signal and interference power increases and the ratio is undetermined.

\begin{figure}[t]
\centering
\includegraphics[width=0.8\linewidth,scale=1.00]{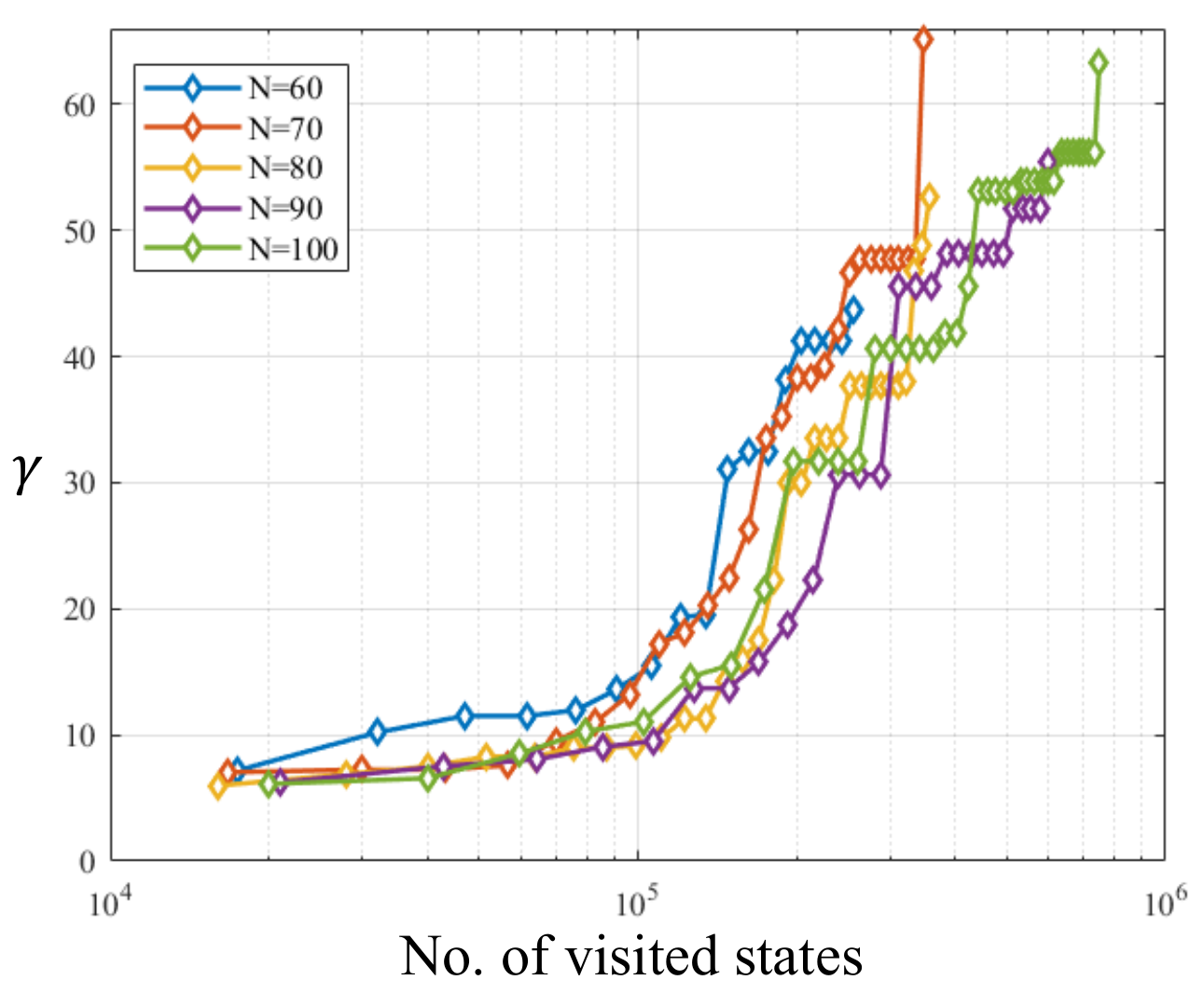}	
\caption{Impact of the different length of the phase codes $N$ on the performance of GASeq.}
\label{fig:change_N1}
\end{figure}

With different code length, the number of visited states of GASeq is shown in Fig.~\ref{fig:change_N1}. When $N=100$, GA discovers $\bm{s}_{\text{GA}}$ with an SCR of $63.23$ after visiting $7.5\times 10^{5}$ states.

\subsection{Impact of initialization, $M$, and $E$}
There are a number of parameters to be determined in GASeq. In this section, we study the impact of these parameters and show how the default parameter settings are chosen.

\begin{figure}[t]
\centering
\includegraphics[width=0.8\linewidth,scale=1.00]{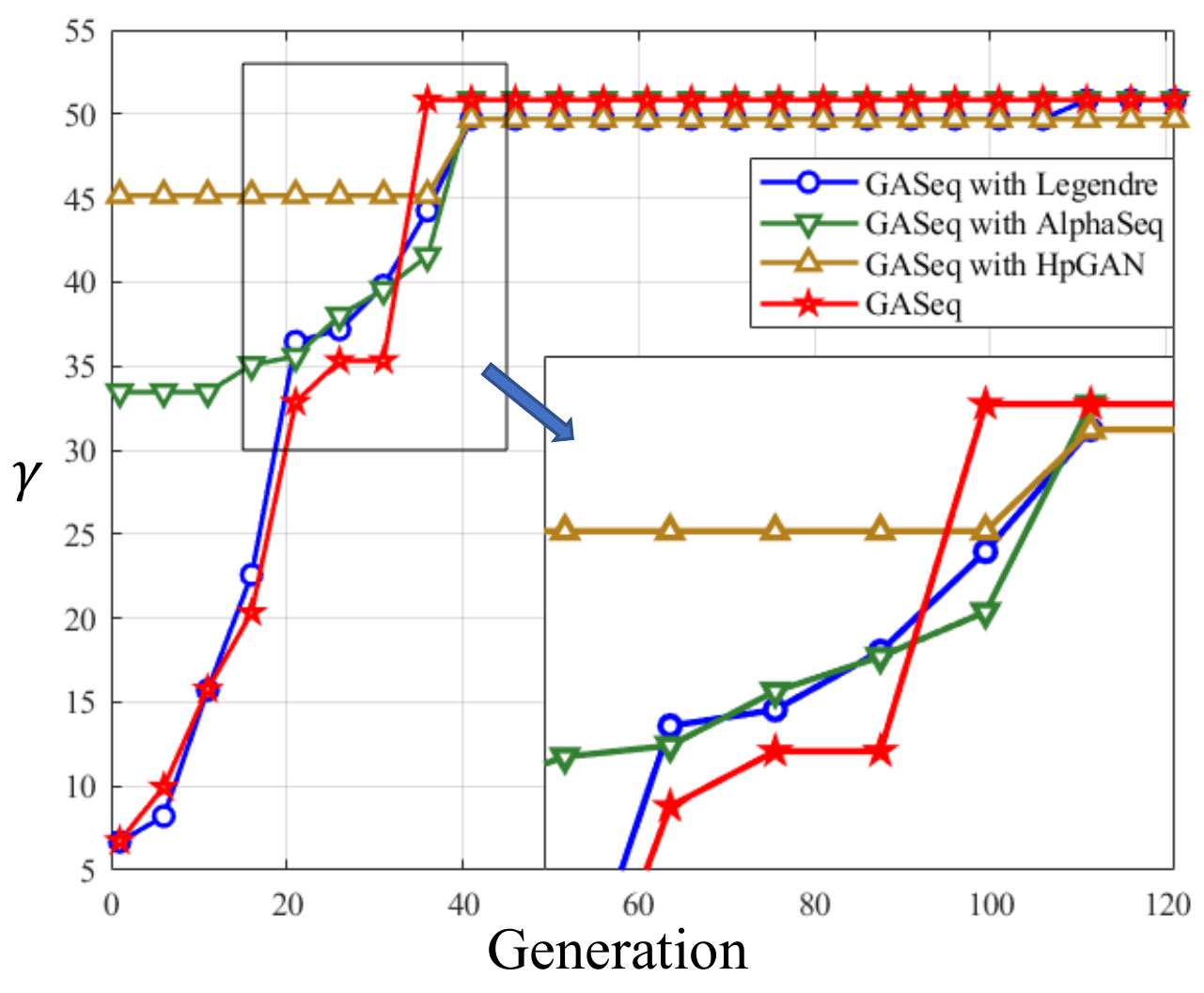}	
\caption{Impact of different initialization schemes on GASeq.}
\label{fig:sim2a}
\end{figure}

In the initialization phase, the first generation of candidates is randomly generated. In addition to that, we can insert existing phase codes in the literature as seeds to generate offspring.
Fig.~\ref{fig:sim2a} compares the evolution of $\gamma$ with different seeds.

As can be seen, adding existing phase codes is actually harmful to the evolution of GASeq. This is not surprising as these ``better'' phase codes dominate the generation, and hence, GASeq explores less compared with the case of pure random initialization.

\begin{figure}[t]
\centering
\includegraphics[width=0.8\linewidth,scale=1.00]{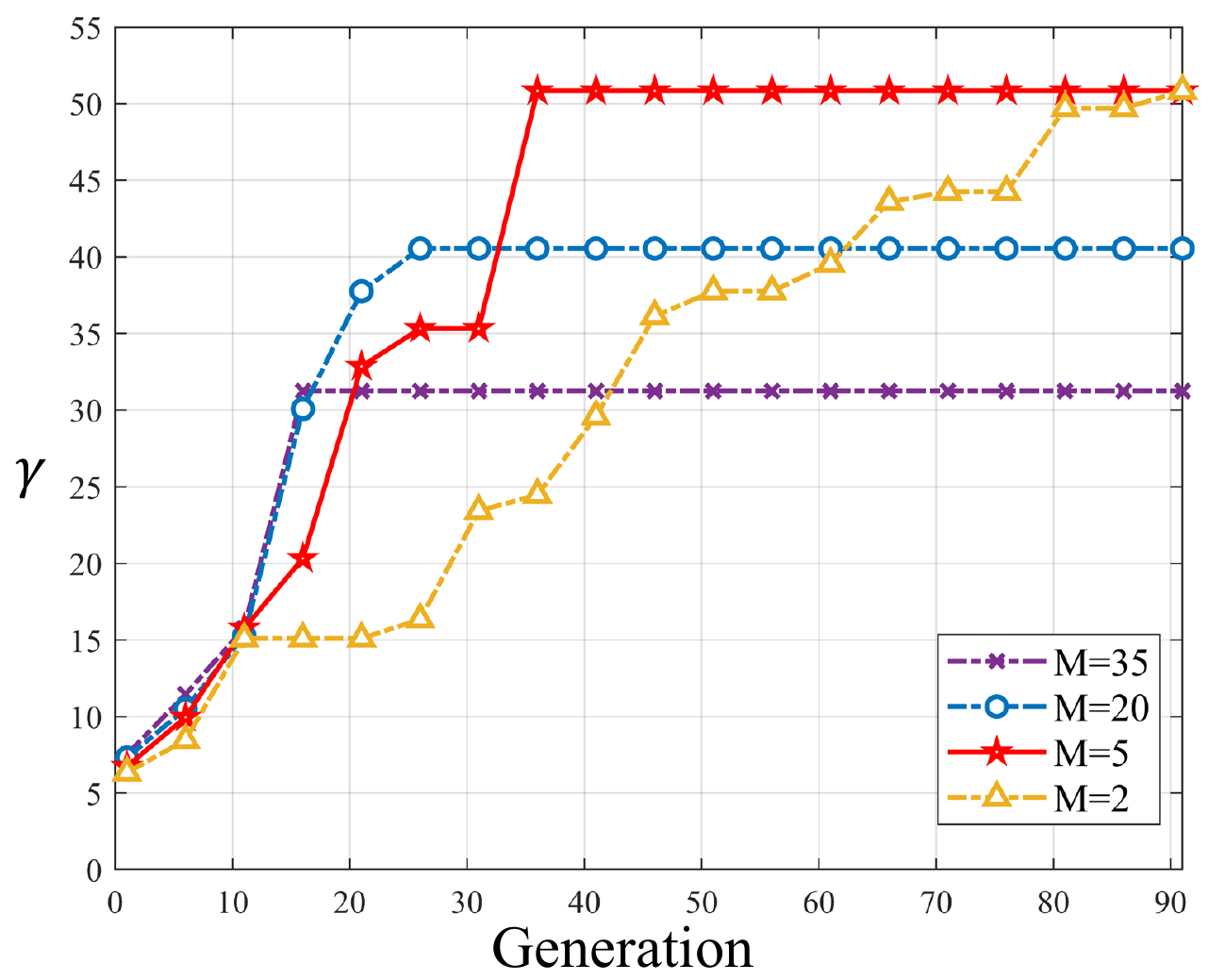}	
\caption{Impact of the number of individuals being selected in each tournament, $M$, on the performance of GASeq.}
\label{fig:sim2b}
\end{figure}

The second important parameter is the number of individuals being selected in each tournament, i.e., $M$, because it directly determines the probability that each candidate goes to the next generation, according to \eqref{eq:probablity}. With different $M$, Fig.~\ref{fig:sim2b} presents the performance of GASeq.

As shown, a large $M$ often leads to suboptimal performance in convergence. This is because the candidates with larger fitness are more likely to go to the next generation if $M$ is large, and hence, the diversity of the population in the next generation decreases.
On the other hand, when $M$ is small, e.g., $M=2$, GASeq can often converge to the same optimum, but the convergence speed can be slow.
Overall, setting $M=5$ is a good choice for GASeq.

\begin{figure}[t]
\centering
\includegraphics[width=0.8\linewidth,scale=1.00]{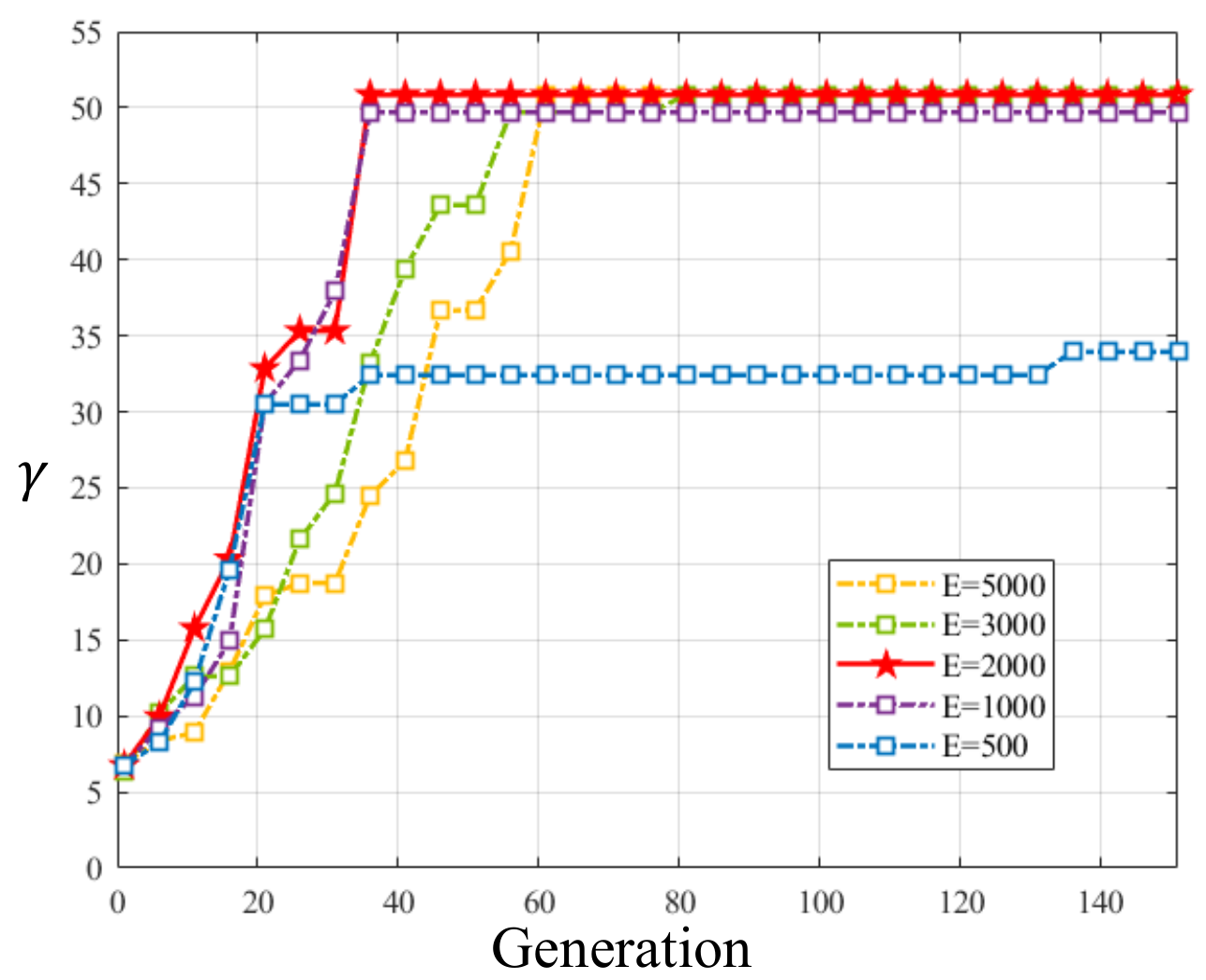}	
\caption{Impact of the elite size $E$ on the performance of GASeq.}
\label{fig:sim2c}
\end{figure}

The final parameter is $E$, the size of the elite set. Succinctly speaking, the setting of $E$ has a similar effect on the performance of GASeq as $M$: a small elite size encourages exploration while a large elite size encourages exploitation. As shown in Fig.~\ref{fig:sim2c}, $E=2000$ is a proper setting as it leads to the optimal phase code while ensuring fast convergence. 

\section{Conclusion} \label{sec:V}
Designing phase codes with low aperiodic autocorrelations is an important  problem in pulse compression radar (PCR) for sidelobe suppression, but is non-trivial due to the lack of mathematical instruments, especially when used in conjunction with the mismatched filter (MMF) receiver to maximize the signal-to-clutter ratio (SCR).
To meet the challenge, this paper put forth a genetic algorithm (GA) approach. Our specific contributions and results are summarized as follows.

We developed GASeq, a GA for phase code discovery in PCR with the MMF receiver.
GASeq exhibits three main advantages over existing deep learning (DL)-based phase codes:
1) Superiority. GASeq discovered better phase codes with higher SCR than the state of the art.
2) Efficiency. GASeq converges much faster and visits significantly fewer states than DL-based approaches to find a sequence with the same SCR.
3) Scalability. GASeq can be readily extended to search longer phase codes, which thwarts DL-based schemes due to the high complexity.

Moving forward, a straightforward extension of GASeq is discovering non-binary (such as polyphase) phase codes to further improve the estimation performance.
To that end, an enhancement on the GA is needed as the phase codes can have continuous, as opposed to discrete, phase or amplitude. A promising candidate for continuous-domain optimization is the differential evolution algorithm and its variants.

\bibliographystyle{IEEEtran}
\bibliography{References}

\end{document}